\documentclass[twocolumn]{aastex631}
\usepackage[normalem]{ulem}  
\usepackage[T1]{fontenc}
\usepackage{ae,aecompl}
\usepackage{graphicx}
\usepackage{amsmath}
\usepackage{amssymb}
\usepackage{supertabular}
\usepackage{hyperref}
\usepackage{xcolor}
\usepackage{academicons}
\usepackage{multirow}
\usepackage{booktabs}
\usepackage{tabularx} 
\usepackage{ragged2e}
\usepackage{threeparttable}

\begin{document}
\title{
A Microphysical Probe of Neutron Star Interiors: Constraining the Equation of State with Glitch Dynamics
}

\author[0000-0001-6836-9339]{Zhonghao Tu}
\affiliation{Department of Astronomy, Xiamen University, Xiamen 361005, People’s Republic of China}

\author[0000-0001-9849-3656]{Ang Li}
\affiliation{Department of Astronomy, Xiamen University, Xiamen 361005, People’s Republic of China}

\email{liang@xmu.edu.cn}

\begin{abstract}

Glitches in neutron stars originate from the sudden transfer of angular momentum between superfluid components and the observable crust. 
By modeling this glitch dynamics--including vortex motion, mutual friction, and angular momentum exchange—one may hope to probe the dense matter equation of state.
In this work, we explore, within a highly idealized three-component framework, whether the glitch rise could in principle carry information about microphysical inputs such as entrainment and mutual friction.
We compute the glitch rise in response to self-consistently calculated microphysical parameters (pinning and mutual friction) based on unified equations of state, and compare theoretical predictions of the overshoot patterns and timing residuals to the 2016 Vela glitch.
Within this specific framework, the models favor crustal superfluid coupling on timescales of order $\sim100$ s, overshoot behavior in the core associated with relatively strong central mutual friction, and rise times consistent with the observed upper limit of 12.6 s.
Using a Markov Chain Monte Carlo analysis of the timing residuals, we then map the regions of parameter space that are compatible with the data under our adopted assumptions. Our analysis favors comparatively weak entrainment in the inner crust and an overall core mutual friction that is weaker than that in the inner crust.
These exploratory results demonstrate that, under such restrictive assumptions, glitch-rise morphology is sensitive to microphysical inputs and that future high-cadence timing observations, interpreted with more realistic dynamical models, could potentially help constrain the internal dynamics and composition of neutron stars.
\end{abstract}


\section{Introduction} \label{sec:intro}
Pulsars, recognized as the most precise clocks in the universe, display sudden increases in the rotational frequency known as glitches \citep{Reichley1969_Nature222-229,Radhakrishnan1969_Nature222-228}. 
These glitches are typically regarded as signatures of superfluidity in stellar interiors \citep{Baym1969_Nature224-872, Anderson1975_Nature256-25}. The phenomenon occurs when neutron vortices undergo catastrophic unpinning, resulting in rapid angular momentum transfer from the superfluid to the observable crust over a short timescale \citep{Alpar1977_ApJ213-527, Haskell2015_IJMPD24-1530008}. 
The accumulated glitch observations has become a valuable tool for investigating pulsar interiors \citep{Alpar1993_ApJ409-345,Chamel2008_LRR11-10,Espinoza2011_MNRAS414-1679,Antonopoulou2022_RPP85-126901, Zhou2022_Universe8-641,Li2025_SCPMA68-119503}. 

Vela (PSR B0833-45) is one of the most frequently glitching pulsars, with 25 observed glitches over the past 56 yr. 
The recent 2016 Vela glitch, with the fractional change in the pulsar frequency $\Delta\nu/\nu=1.431\times10^{-6}$, provided a unique opportunity to study pulse-to-pulse glitch dynamics \citep{Palfreyman2018_Nature556-219}. 
\citet{Ashton2019_NA36-844}, using a phenomenological timing model, identified an overshoot in the inferred crustal frequency evolution and placed an upper limit of 12.6 s on the glitch rise time.
Within a three-component dynamical framework, \citet{Graber2018_ApJ865-23} reproduced the observed overshoot and relaxation by adopting density-dependent mutual friction coefficients in the crust, and showed that the modeled rise behavior depends on the relative strengths of crustal and core mutual friction.
\citet{Pizzochero2020_A&A636-A101} further examined the relative contributions of crustal and core superfluid components to the same event, explicitly including entrainment effects in their model.
Using a Bayesian inference approach, \citet{Montoli2020_A&A642-A223} confirmed the presence of an overshoot in the timing residuals and constrained the glitch rise time to be shorter than $\sim 6$ s within their adopted model.
Despite these advances, it remains unclear how specific microscopic inputs (e.g., pinning energies, entrainment strength, and mutual friction mechanisms) affect the glitch rise, and it is not yet established to what degree current observations can uniquely constrain these microphysical parameters.

In this study, we simulate the detailed coupling dynamics of the glitch rise with the self-consistently calculated microphysical inputs, perform Markov Chain Monte Carlo (MCMC) analyses of the timing residuals of the 2016 Vela glitch to examine whether glitch rise can encode information of microscopic physics.

The structure of the paper is as follows. Section \ref{subsec:glitchmodel} introduces the extended three-component glitch model. Section \ref{sec:inputs} describes the microphysical inputs adopted in this work: Sections \ref{subsec:EoS}–\ref{subsec:friction} outline the theoretical framework for constructing a unified equation of state (EOS), the analytical representation of the entrainment effect, and the determination of dimensionless drag parameters in both the crust and core, respectively. Section \ref{sec:mutualfriction} presents the results of mutual friction coefficients both for crustal and core superfluids. Theoretical predictions for the glitch rise are presented in Section \ref{sec:glitchrise}: the differential rotation of the internal superfluid is shown in Section \ref{subsec:differerntialtotation}, while the evolution of the crustal frequency and corresponding model residuals are discussed in Section \ref{subsec:crustalfrequency}; Section \ref{subsec:effects} examines the effects of the symmetry energy slope $L_0$, entrainment, and neutron star mass on the glitch rise. In Section \ref{sec:Vela2016}, the three-component model is applied to the 2016 Vela glitch: Section \ref{subsec:mcmc} presents the MCMC simulations and parameter correlations, and Section \ref{subsec:constrain to vela} discusses the resulting constraints on the Vela's mass, EOS, and interiors. A summary of the main conclusions is provided in Section \ref{sec:summary}.

\section{Three-component glitch model}\label{subsec:glitchmodel}

The multi-fluid simulation of a pulsar glitch involves a computationally challenging 3D problem. Following the work of \cite{Antonelli2016_MNRAS464-721}, under the assumptions of an axisymmetric star and straight rigid vortices, the 3D simulation can be reduced to a simpler 1D problem by projecting the spherical star onto its equatorial plane. The minimal model capable of producing a glitch rise accompanied by an overshoot is the three-component model, which consists of two distinct, non-overlapping neutron superfluid components and a ``crust" component that includes the solid crust and all components tightly coupled to solid crust. 
We emphasize that the assumption of straight, rigid vortices is a working, geometrical simplification common to cylindrical glitch models, introduced to render the problem tractable rather than as a realistic description of vortex configurations in neutron star interiors (see discussion in \citealt{Antonelli2022_NSPhysicsGlitches}). This rigid-vortex approximation should therefore be viewed as part of the modelling framework and borne in mind when interpreting the quantitative results.

Within this framework, the superfluid is threaded by an array of such vortices aligned with the rotation axis.
Using the standard cylindrical coordinates $(x,~ \theta,~z)$, microphysical inputs, such as the pinning force per unit length $f_{\rm{p}}$ and the drag parameter $\mathcal{R}$, are integrated along the $z$-axis at a given cylindrical radius $x$ to describe the 1D collective dynamical behavior of straight, rigid vortices. If the vortices are assumed to be discontinuous across the crust–core interface, the superfluid can be decomposed into a crustal superfluid with the momentum of inertia (MoI) $I_{\text{sf}}^{\text{cr}}$ and a core superfluid with MoI $I_{\text{sf}}^{\text{co}}$. While vortices can be pinned either to the nuclear lattice in the inner crust or to flux tubes in the outer core, our three-component model, for simplicity, restricts itself to the nuclear lattice pinning scenario, in which the crustal superfluid acts as an angular momentum reservoir.

From the conservation of vorticity and the total angular momentum balance, the macroscopic evolution equations for the angular velocities of the crustal superfluid $\Omega_{\rm{sf}}^{\rm{cr}}$, the core superfluid $\Omega_{\rm{sf}}^{\rm{co}}$, and the ``crust'' component $\Omega_{\rm{p}}$ are derived as follows: 
\begin{widetext}
\begin{eqnarray}
\dot{\Omega}_{\rm{sf}}^{\rm{cr}} =& \mathcal{B}_{\rm{cr}}(x) \left[ 2\Omega_{\rm{sf}}^{\rm{cr}} + x \frac{\partial \Omega_{\rm{sf}}^{\rm{cr}}}{\partial x} \right] (\Omega_{\rm{p}} - \Omega_{\rm{sf}}^{\rm{cr}}), \label{eq:Omegadot_sf}\\
\dot{\Omega}_{\rm{sf}}^{\rm{co}} =& \mathcal{B}_{\rm{co}}(x) \left[ 2\Omega_{\rm{sf}}^{\rm{co}} + x \frac{\partial \Omega_{\rm{sf}}^{\rm{co}}}{\partial x} \right] (\Omega_{\rm{p}} - \Omega_{\rm{sf}}^{\rm{co}}), \\
I_{\rm{p}}\dot{\Omega}_{\rm{p}} =&  - 2\pi\int_0^{R_{\rm{oi}}} b_{\rm{cr}}(x) x^3 \dot{\Omega}_{\rm{sf}}^{\rm{cr}}\, dx - 2\pi\int_0^{R_{\rm{cc}}} b_{\rm{co}}(x) x^3 \dot{\Omega}_{\rm{sf}}^{\rm{co}} \, dx, \label{eq:Omegadot_crust}
\end{eqnarray}
\end{widetext}
where we have neglected the external spin-down torque $N_{\rm{ext}}$. $I_{\rm{p}}$ is the MoI of   ``crust'' component. $R_{\rm{oi}}$ and $R_{\rm{cc}}$ are the radius corresponding to the outer-inner crust and crust-core interfaces, respectively. The dimensionless mutual friction coefficients $\mathcal{B}_{\rm{cr}}(x)$ for the crustal superfluid and $\mathcal{B}_{\rm{co}}(x)$ for the core superfluid are obtained by the balance of the integrated drag force and Magnus force: 
\begin{align}
    \mathcal{B}_{\rm{cr}}(x)=\frac{b_{\rm{cr}}(x)d_{\rm{cr}}(x)}{d_{\rm{cr}}^2(x)+a_{\rm{cr}}^2(x)}, \label{eq:Bcr}\\
    \mathcal{B}_{\rm{co}}(x)=\frac{b_{\rm{co}}(x)d_{\rm{co}}(x)}{d_{\rm{co}}^2(x)+a_{\rm{co}}^2(x)}, \label{eq:Bco}
\end{align}
where $\rho_{\rm{n}}$ is the neutron superfluid density. The restricted line integrals are $a_{\rm{cr}}(x)=2\int_{z_{\rm{cr}}^-(x)}^{z_{\rm{cr}}^+(x)}\rho_{\mathrm{n}}(r)dz$, $b_{\rm{cr}}(x)=2\int_{z_{\rm{cr}}^-(x)}^{z_{\rm{cr}}^+(x)}\frac{\rho_{\mathrm{n}}(r)}{1-\epsilon_{\rm{n}}(r)}dz$, and $d_{\rm{cr}}(x)=2\int_{z_{\rm{cr}}^-(x)}^{z_{\rm{cr}}^+(x)}\rho_{\mathrm{n}}(r)\mathcal{R_{\rm{cr}}}(r)dz$ with $z_{\rm{cr}}^+(x) = \sqrt{R_{\rm{oi}}^2-x^2}$ and $z_{\rm{cr}}^-(x) = \sqrt{R_{\rm{cc}}^2-x^2}$ ($z_{\rm{cr}}^-(x)=0$ if $x>R_{\rm{cc}}$) for the crustal superfluid; $a_{\rm{co}}(x)=2\int_0^{z_{\rm{co}}^+(x)}\rho_{\mathrm{n}}(r)dz$, $b_{\rm{cr}}(x)=2\int_0^{z_{\rm{co}}^+(x)}\frac{\rho_{\mathrm{n}}(r)}{1-\epsilon_{\rm{n}}(r)}dz$, and $d_{\rm{cr}}(x)=2\int_0^{z_{\rm{co}}^+(x)}\rho_{\mathrm{n}}(r)\mathcal{R_{\rm{co}}}(r)dz$ with $z_{\rm{co}}^+(x) = \sqrt{R_{\rm{cc}}^2-x^2}$ for the core superfluid. $\mathcal{R_{\rm{cr}}}$ and $\mathcal{R_{\rm{co}}}$ are the dimensionless drag parameters for the crust and core, respectively. Entrainment enters the above equations through the relation between the momentum per particle $\boldsymbol{p}_n$ at position $\boldsymbol{x}$ and the local kinematic velocities of the two components $\boldsymbol{v}_{\rm{n}}$ and $\boldsymbol{v}_{\rm{p}}$: $\boldsymbol{p}_n = m_{\rm{n}}[(1-\epsilon_{\rm{n}})\boldsymbol{v}_{\rm{n}} + \epsilon_{\rm{n}} \boldsymbol{v}_{\rm{p}}]$, where $\epsilon_{\rm{n}}$ is the entrainment parameter. These microphysical quantities $\rho_{\rm{n}}$, $\epsilon_{\rm{n}}$, and $\mathcal{R}$ are functions of spherical radius $r$, where $r=\sqrt{x^2+z^2}$. With the above definitions, $I_{\rm{sf}}^{\rm{cr}}$ and $I_{\rm{sf}}^{\rm{co}}$ can be written as:
\begin{align}
    I_{\rm{sf}}^{\rm{cr}}=2\pi\int_{z_{\rm{cr}}^-(x)}^{z_{\rm{cr}}^+(x)} x^3b_{\rm{cr}}(x) \, dr ,\\
    I_{\rm{sf}}^{\rm{co}}=2\pi\int_0^{z_{\rm{co}}^+(x)} x^3b_{\rm{co}}(x)\, dr.
\end{align} 
The MoI of ``crust'' component can be calculated by $I_{\rm{p}}=I - I_{\rm{sf}}^{\rm{cr}}-I_{\rm{sf}}^{\rm{co}}$. The total MoI $I$ of stellar is $I=\int_0^Rr^4\rho\,dr$ with the stellar radius $R$ and the mass density $\rho$.

The initial conditions for solving Eqs. (\ref{eq:Omegadot_sf})--(\ref{eq:Omegadot_crust}) are taken from the typical values of Vela, i.e., $\Omega_{\rm{p}}(0)\approx\Omega_{\rm{sf}}^{\rm{co}}(x,0)\approx70.34026$ rad/s ($\nu\approx11.195$ Hz) and $\Omega_{\rm{sf}}^{\rm{cr}}(x,0)-\Omega_{\rm{p}}(0)\approx6.3\times10^{-3}$ rad/s \citep{Palfreyman2018_Nature556-219,Graber2018_ApJ865-23}. Note that the initial lag between the crustal superfluid and the ``crust'' component is assumed to be spatially uniform, so that the entire crustal superfluid acts as an angular momentum reservoir. We take the boundary conditions $\dot{\Omega}_{\rm{sf}}^{\rm{cr}}=\dot{\Omega}_{\rm{sf}}^{\rm{co}}=0$ at $x=0$ and $x=R$.

\section{Microphysical Inputs}\label{sec:inputs}

To build a self-consistent glitch model, we first identify the effective dynamical quantities required by the three-component model: the mutual friction coefficients in the crustal and core superfluids, $\mathcal{B}_{\rm cr}(x)$ and $\mathcal{B}_{\rm co}(x) $; the MoI of the crustal and core superfluids, $I_{\rm sf}^{\rm cr}$ and $I_{\rm sf}^{\rm co}$, and the MoI of the ``crust'' component, $I_{\rm p}$; the line integrals for the superfluid density after considering entrainment, $b_{\rm cr}(x)$ and $b_{\rm co}(x)$. To compute them, we require  several microphysical inputs, which are detailed in the following three sections. First, we construct a unified EOS for the neutron star crust and core, which provides the stellar structure and composition. From this unified EOS, we also self-consistently derive the pinning energy in the inner crust following \citet{Tu2025_ApJ984-200}. Second, we determine the entrainment parameters in both the crust and core. Third, we calculate the dimensionless drag parameters. With these inputs, we then derive the mutual friction coefficients and the remaining dynamical quantities in a self-consistent manner.

\subsection{Unified description for the crust and core of neutron stars}\label{subsec:EoS}

The EOS determines the stellar structure and is therefore crucial for the superfluid coupling dynamics of a glitch. Unified EOS demands that all regions of a neutron star—such as the crust and core—be computed within a single, self-consistent nuclear many-body framework, employing the same underlying nuclear interaction and ensuring consistent composition across different layers~\citep[e.g.,][]{2013A&A...559A.128F}.
Based on the relativistic mean field (RMF) model \citep{Fetter1972_PT25-54,Walecka1974_APNY83-491,Serot1992_RPP55-1855}, we construct the unified EOS using two sets of density-dependent effective interaction: DD-ME2 \citep{Lalazissis2005_PRC71-024312} and PKDD \citep{Long2004_PRC69-034319}. The unified EOS within the RMF framework was detailed in our previous studies \citep[e.g.,][]{Tu2025_ApJ984-200}, to which readers are referred for further details.

\begin{table}[htbp]
  \centering
  \caption{
  Nuclear matter saturation properties for the DD-ME2 and PKDD effective interactions. Listed are the saturation density~$\rho_{0}$~(fm$^{-3}$), binding energy per particle $E/A$ (MeV), incompressibility $K_0$ (MeV), skewness $Q_0$ (MeV), symmetry energy $J_0$ (MeV), slope of symmetry energy $L_0$ (MeV), and Dirac effective mass of neutron $M^{\star}/M$.}\label{tab:satpros}
  \setlength\tabcolsep{2.5pt}
  \begin{tabular}{lccccccc}
  \hline
  \hline
         & $\rho_{0}$ & $E/A$ & $K_0$ & $Q_0$ & $J_0$ & $L_0$ & $M_n^{\star}/M_n$ \\
                          & $\rm{fm}^{-3}$ & MeV & MeV & MeV & MeV & MeV \\
  \hline
    DD-ME2 & 0.152 & $-$16.14 & 251.1 & 479 & 32.30 & 51.26 & 0.572 \\
    PKDD    & 0.150 & $-$16.27 & 262.2 & $-$115 & 36.86 & 90.21 & 0.570 \\
  \hline
  \end{tabular}
\end{table}

\begin{table*}[htbp]
  \centering
  \caption{Coupling parameters $g_{\rho}$ and $a_{\rho}$ between nucleons and $\rho$ meson, radius $R_{\rm{cc}}$ of neutron star core, thickness $d$ of the inner crust, stellar radius $R$, and predicted maximum mass $M_{\rm{TOV}}$ for the DD-ME2 and PKDD interactions, adjusted to different values of the symmetry energy slope $L_0$. $R_{cc,1.4}$, and $R_{1.4}$ correspond to values for an neutron star with $1.4M_{\odot}$.}\label{tab:vectorcouple}
  \setlength\tabcolsep{15pt}
  \begin{tabular}{lrrrrrrrr}
    \hline
    \hline
     & \multicolumn{4}{c}{DD-ME2} & \multicolumn{4}{c}{PKDD} \\
     \cmidrule(r){2-5} \cmidrule(r){6-9}
     $L_0$ (MeV) & 30 & 40 & 60 & 80 & 30 & 40 & 60 & 80 \\
    \hline
    $g_{\rho}$  & 3.357 & 3.528 & 3.792 & 4.010 & 3.629 & 3.769 & 4.006 & 4.206 \\
    $a_{\rho}$  & 0.891 & 0.721 & 0.460 & 0.258 & 0.827 & 0.684 & 0.453 & 0.267 \\
    $R_{\rm{cc,1.4}}$ (km) & 11.822 & 11.844 & 12.081 & 12.469 & 11.644 & 11.672 & 11.840 & 12.143 \\
    $R_{1.4}$ (km) & 13.011 & 13.087 & 13.296 & 13.591 & 12.876 & 12.944 & 13.149 & 13.454 \\
    $M_{\rm{TOV}}$ ($M_{\odot}$) & 2.497 & 2.491 & 2.477 & 2.468 & 2.356 & 2.350 & 2.336 & 2.325 \\
    \hline
  \end{tabular}
\end{table*}

The original DD-ME2 and PKDD demonstrate support for the existence of massive neutron stars ($>2M_\odot$). The extension of DD-ME and PKDD in the isospin vector channel can be obtained by adjusting the density-dependent coupling parameters, i.e., $g_{\rho}$ and $a_{\rho}$, for the $\rho$ meson; see \citep{Tu2025_ApJ984-200} for details. In Table \ref{tab:satpros}, we list the saturation properties of nuclear matter 
for the original effective interactions. The coupling parameters of extensions of two effective interactions are listed in Table \ref{tab:vectorcouple} with $L_0=30,~40,~60$, and 80 MeV.  These coupling parameters are obtained by fixing the symmetry energy $E_{\rm{sym}}$ at baryon number density 
$n_{\mathrm{B}}=0.11$ fm$^{-3}$ but adjusting the symmetry energy slope $L_0$ at the saturation density.  

Based on the unified EOSs, the nonuniform structure of the inner crust at a given density can be obtained self-consistently, which in turn allows for the self-consistent calculation of the microscopic pinning energy and the mesoscopic pinning force. Readers interested in the details are referred to \citet{Tu2025_ApJ984-200}.

\subsection{Entrainment}\label{subsec:entrainment}
In the inner crust, the entrainment effect causes a part of free neutrons to flow with the nuclei, resulting in a reduction of the superfluid density. We use the notation of the effective mass $m_{\mathrm{n}}^*$ to characterize the entrainment effect, the entrainment parameter $\epsilon_{\mathrm{n}}=1-m_{\mathrm{n}}^*/m_{\mathrm{n}}$ with the bare neutron mass $m_{\mathrm{n}}=M_{\mathrm{n}}$. In the framework of the band theory of solids, \citet{Chamel2012_PRC85-035801} calculated the effective mass at different densities within the inner crust and found that it is density-dependent. $m_{\mathrm{n}}^*/m_{\mathrm{n}}$ is 1 at the transition density $n_{\mathrm{oi}}$ between the outer and inner crust.  $m_{\mathrm{n}}^*/m_{\mathrm{n}}$ increases monotonically with density, peaks at approximately 13.6 around $n_B = 0.03\,\text{fm}^{-3}$, and then decreases monotonically as density increases, reaching approximately 1.0 at the transition density $n_{\mathrm{cc}}$ between the crust and the core. This indicates a strong entrainment effect in the inner crust. However, more studies indicated that the strength of the entrainment effect remains a matter of debate \citep{Oyamatsu1994_NPA578-181,Carter2005_NPA748-675,Chamel2005_NPA747-109,Kashiwaba2019_PRC100-035804, 2025PhRvC.111d5803C,2025PhRvL.135m2701A}. 

To simplify the treatment of entrainment and its density dependence in this work, we fix $m_{\mathrm{n}}^*/m_{\mathrm{n}}$ at two boundary densities $n_{\mathrm{oi}}$ and $n_{\mathrm{cc}}$ to 1, fix the peak density of $m_{\mathrm{n}}^*/m_{\mathrm{n}}$ to $0.03\ \mathrm{fm}^{-3}$, and treat the peak value $(m_{\mathrm{n}}^*/m_{\mathrm{n}})_{\mathrm{peak}}$ as a free adjustable parameter. The values of $m_{\mathrm{n}}^*/m_{\mathrm{n}}$ at other densities are obtained via linear interpolation between the peak point and the two boundary points. Neutron entrainment in the core are not taken into account in this work.

\subsection{Dimensionless drag parameter}\label{subsec:friction}

During a glitch, there exists a local velocity $\Delta v$ between a vortex and a nucleus, then the dynamics of the vortex is fully characterized by a dimensionless drag parameter $\mathcal{R}$. In the inner crust, when $\Delta v$ is sufficiently large, the dissipation is dominated by Kelvin wave excitation. In this study, we discuss Kelvin wave dynamics based on the work of \citet{Epstein1992_ApJ387-276}. The drag parameter can be expressed as:
\begin{eqnarray}\label{eq:R_EBJ}
\mathcal{R}_{\mathrm{EB}}\simeq1.4\left(\frac{\mu}{\hbar}\right)^{1/2}\left(\frac{E_{\mathrm{p}}\delta}{\rho_{\mathrm{n}}\kappa}\right)^{1/2}\frac{R_{\mathrm{d}}}{a^{3/2}}.
\end{eqnarray}
Here $\mu$ is the effective mass of the neutrons associated with the vortex tension; we follow \citet{Graber2018_ApJ865-23} and set $\mu\approx M_n/2$ with the nucleon mass $M = 939$ MeV. $\kappa$ is quantum of circulation with a approximate value $2.0\times10^{-3}~\mathrm{cm}^2~\mathrm{s}^{-1}$. $R_{\rm{d}}$ and $a = 4R_{\mathrm{WS}}/\sqrt{3}$ are the radius of nuclear cluster and bcc lattice constant, which are both obtained in Section \ref{subsec:EoS}. The effective pinning energy $E_{\rm{p}}\delta$ is defined by the relation $f_{\rm{pin}} = E_{\rm{p}}\delta / \ell a$ with $\ell\approx R_{\rm{WS}}$, The pinning energy $E_{\rm{p}}$ is calculated using the unified EOSs from Section \ref{subsec:EoS} and the semi-classical method detailed in \citet{Tu2025_ApJ984-200}. The mesoscopic pinning force $f_{\rm{pin}}$ can be estimated by the method developed in \citet{Seveso2016_MNRAS455-3952}, with a long vortex rigidity length $l =5000R_{\rm{WS}}$.

In the core, we focus mainly on the vortex dynamics which is dominated by the scattering of electrons off magnetized vortices. The mutual friction coefficient reads \citep{Andersson2006_MNRAS368-162}:
\begin{widetext}
\begin{equation}\label{eq:friction_core}
    \mathcal{R}_{\mathrm{core}}\simeq4\times10^{-4}\left(\frac{\delta \mathcal{M}_{p}^*}{\mathcal{M}_p}\right)^2\left(\frac{\mathcal{M}_p}{\mathcal{M}_p^*}\right)^{1/2}\left(\frac{Y_p}{0.05}\right)^{7/6}\left(\frac{\rho}{10^{14}~\text{g/cm}^3}\right)^{1/6},
\end{equation}
\end{widetext}
where $\mathcal{M}_p=M_p$ is the bare proton mass, while $\mathcal{M}_{p}^*$ is the effective proton mass associated with the proton entrainment effect. $\delta\mathcal{M}_{p}^* = \mathcal{M}_p-\mathcal{M}_p^*$. $Y_p$ is the proton fraction and $\rho=m_un_B$ is the mass density in the unit of g/cm$^3$. To fall within the weak coupling regime, an effective proton mass $\mathcal{M}_p^*/\mathcal{M}_p=0.5$--$0.7$ is recommended. An alternative and stronger source of mutual friction ($\sim10^{-2}$) in the core may originate from the pinning of neutron vortices to fluxtubes \citep{Link2003_PRL91-101101,Sidery2009_MNRAS400-1859,Graber2018_ApJ865-23}, provided that core protons form a type II superconductor \citep{Baym1969_Nature224-673}.
The stark contrast between these two mechanisms is summarized in Table \ref{tab:friction_comparison}. The orders-of-magnitude difference in $\mathcal{B}_{\rm core}$ provides a discriminant. 

\section{Mutual friction}\label{sec:mutualfriction}

\begin{figure}
\centering
\includegraphics[width=0.45\textwidth]{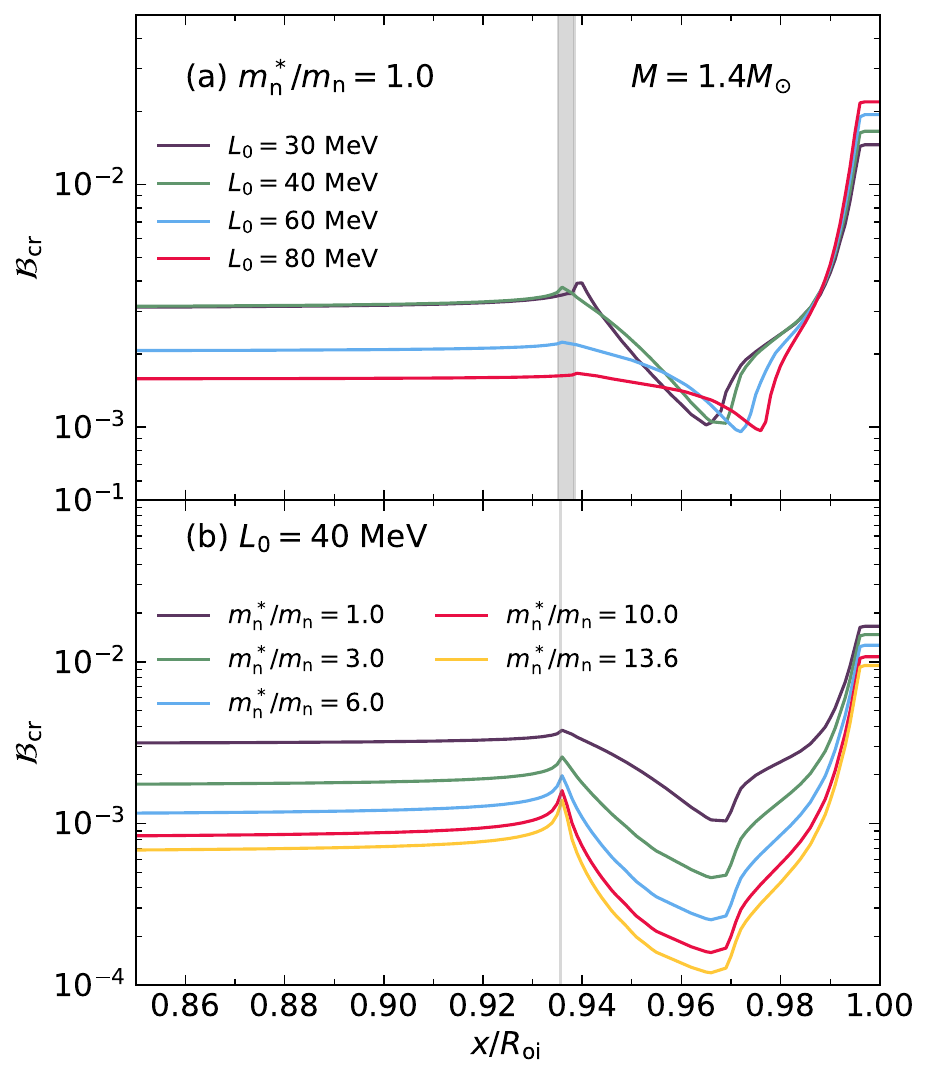}
\caption{Mutual friction coefficient $\mathcal{B}_\mathrm{cr}$ as a function of normalized cylindrical radius $x/R_{\rm{oi}}$ for a $1.4M_{\odot}$ neutron star. 
(a) Dependence on the symmetry energy slope $L_0$ for the PKDD interaction, with no entrainment ($m_{\mathrm{n}}^*/m_{\mathrm{n}}=1.0$). (b) Dependence on the effective mass $m_{\mathrm{n}}^*/m_{\mathrm{n}}$ for PKDD with $L_0=40$ MeV. The gray-shaded region and the gray solid vertical line mark the radial positions of the crust–core interface in the equatorial plane. Stronger entrainment (larger $m_{\mathrm{n}}^*/m_{\mathrm{n}}$) significantly suppresses the crustal mutual friction coefficients. 
}
\label{fig:friction_crust}
\end{figure}

\begin{table*}[htbp]
  \centering
  \caption{Comparison of Friction Mechanisms in Neutron Stars.
  }
  \label{tab:friction_comparison}
\begin{tabularx}{\textwidth}{
    >{\RaggedRight}p{3.5cm}
    >{\RaggedRight}p{4.0cm}
    >{\RaggedRight}p{4.0cm}
    >{\RaggedRight}p{5.0cm}
}
\toprule
 \textbf{Mechanism} & \textbf{Assumptions} & \textbf{Observational Signatures} & \textbf{Parameters and Ranges} \\
\midrule
Crust: Kelvin Wave Excitation &
Rigid vortices ($l \approx 5000 R_{\text{ws}}$); \newline
Semiclassical $E_{\rm{p}}$ calculation; \newline
BCC nuclear lattice; \newline
Global entrainment effect. &
Rise time $< 12.6$ s (Vela); \newline
Crustal frequency overshoot (if relatively weak core friction); \newline
Complex relaxation profile &
Mutual friction coefficient $\mathcal{B}_{\text{cr}}\sim10^{-4}$ to $10^{-1}$; \newline
Entrainment effect $m_{\mathrm{n}}^*/m_{\mathrm{n}}$: 1.0-13.6 ; \newline
Symmetry energy slope $L_0$: 30-80 MeV. \\
\midrule
Core: Electron Scattering &
Proton superconductivity; \newline
Magnetized vortices; \newline
Weak coupling regime. &
Overshoot decay time $\sim$100 s; \newline
Core superfluid frequency overshoot; \newline
Long-term momentum transfer. &
Mutual friction coefficient $\mathcal{B}_{\text{co}}\sim10^{-4}$; \newline
Effective mass ratio $\mathcal{M}^*_p/\mathcal{M}_p$: 0.5--0.7; \newline
Proton fraction $Y_p\sim$ 0.05--0.1, core density $\rho\sim10^{14}$ g/cm$^3$. \\
\midrule
Core: Vortex-Fluxtube Pinning (Alternative) &
Type-II superconductor;
\newline
Strong vortex-fluxtube interaction; \newline
Magnetic interaction dominates. \newline &
Stronger core-``crust'' coupling; \newline
Alters rise/overshoot (not observed). &
$\mathcal{B}_{\text{co}}\sim 10^{-2}$ (if dominant); \newline
Pinning energy $E_{\rm{p}}\sim5$ MeV,
Magnetic field $B\sim10^{12}$ G, London penetration depth $\lambda\sim$ 100 fm \citep{Graber2018_ApJ865-23}. \\
\bottomrule
\end{tabularx}
\end{table*}

\begin{figure}
\centering
\includegraphics[width=0.45\textwidth]{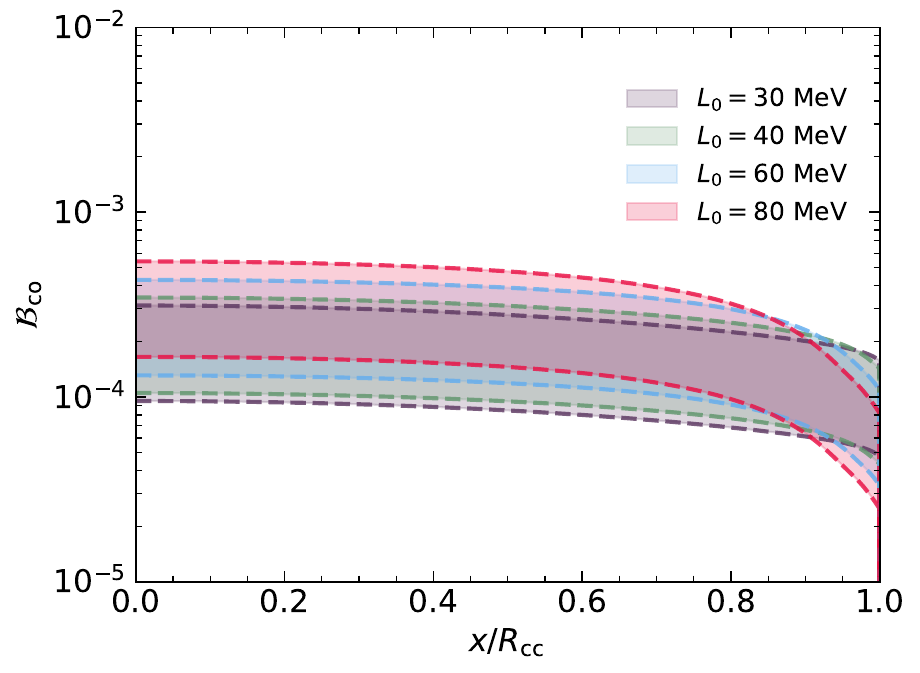}
\caption{Mutual friction coefficients $\mathcal{B}_{\mathrm{co}}$ as a function of normalized cylindrical radius $x/R_{\mathrm{cc}}$ for PKDD with different $L_0$ and fixed $1.4M_{\odot}$ neutron star mass. The shaded region indicates the range of $\mathcal{B}_{\mathrm{co}}$ for  $\mathcal{M}_{p}^*/\mathcal{M}_{p} = 0.5$--$0.7$, with the upper and lower boundaries corresponding to $\mathcal{M}_{p}^*/\mathcal{M}_{p} =0.5$ and $\mathcal{M}_{p}^*/\mathcal{M}_{p} = 0.7$, respectively.
}
\label{fig:friction_core}
\end{figure}

A necessary condition for the occurrence of an overshoot during the glitch rise is that $\mathcal{B}_{\rm{cr}}$ be sufficiently larger than $\mathcal{B}_{\rm{co}}$ overall, so that the excess angular momentum can be temporarily stored in the crust. With the approximation and microphysical inputs given in Section \ref{sec:inputs}, the mutual friction coefficients follow from Eqs. (\ref{eq:Bcr})--(\ref{eq:Bco}). We show the mutual friction coefficients $\mathcal{B}_{\rm{cr}}$ ($\mathcal{B}_{\rm{co}}$) as a function of the normalized cylindrical radius $x/R_{\text{oi}}$ ($x/R_{\text{cc}}$), as we can see in Figure \ref{fig:friction_crust} (\ref{fig:friction_core}). Here we calculate mutual friction coefficients only at some discrete densities, we apply a spline function to interpolate mutual friction coefficient between the maximum and minimum densities. For densities outside this range, we keep mutual friction coefficient constant for simplicity. 

From Figure \ref{fig:friction_crust}(a), for a given EOS and entrainment parameter (hereafter we denote $(m_{\mathrm{n}}^*/m_{\mathrm{n}})_{\rm{peak}}$ as $m_{\mathrm{n}}^*/m_{\mathrm{n}}$), $\mathcal{B}_{\mathrm{cr}}$ is nearly constant for $x < 0.85 R_{\mathrm{oi}}$, with values showing only minor variation for $L_0 \lesssim 40$ MeV, but decreasing as $L_0$ increases from 40 to 80 MeV. This effect primarily arises from the reduction of the pinning energy caused by a high symmetry energy slope, see \citet{Tu2025_ApJ984-200}. For $R_{\rm{cc}} < x < R_{\rm{oi}}$, $\mathcal{B}_{\rm{cr}}$ first decreases and then increases with increasing $x$. The minimum value is insensitive to $L_0$, while the radius at which the minimum occurs increases with $L_0$, consistent with the dependence of the pinning configuration transition density on $L_0$ \citep{Tu2025_ApJ984-200}. $\mathcal{B}_{\mathrm{cr}}$ falls within the range of $1\times10^{-3}$ to $1\times10^{-2}$. Approaching the equator ($x\approx R_{\rm{oi}}$) of stellar, $\mathcal{B}_{\mathrm{cr}}$ is on the order of $10^{-2}$; in other regions of $R_{\rm{cc}} < x < R_{\rm{oi}}$, $\mathcal{B}_{\mathrm{cr}}$ remains on the order of $10^{-3}$. This indicates that the equator region is a strong coupling region, while the region near the minimum is the weak coupling region. Since the superfluid density increases with crustal depth (or increasing baryon number density), the region near the inner-outer crust interface carries a minority of the superfluid's mass. This results in the strongly coupling region contributing little to angular momentum transfer during a glitch. From Figure \ref{fig:friction_crust}(b), we see that the entrainment effect leads to a monotonic reduction of the mutual friction coefficient for the crustal superfluid, indicating that stronger entrainment more effectively suppresses the angular momentum transfer rate during glitch rise. The results obtained with DD-ME2 show qualitatively the same trend.

In Figure \ref{fig:friction_core}, we show the mutual friction coefficients $\mathcal{B}_{\mathrm{co}}$ as a function of normalized cylindrical radius $x/R_{\mathrm{cc}}$ in the core. We can see that $\mathcal{B}_{\mathrm{co}}$ exhibits monotonic growth toward the neutron star center before reaching a plateau, with its magnitude extending from $10^{-5}$ to $10^{-4}$. $\mathcal{B}_{\mathrm{co}}$ is generally smaller than $\mathcal{B}_{\mathrm{cr}}$, satisfying the condition for overshoot \citep{Antonelli2022_NSPhysicsGlitches}. A smaller proton effective mass leads to a larger $\mathcal{B}_{\mathrm{co}}$. Near $x=R_{\rm{cc}}$, $\mathcal{B}_{\mathrm{co}}$ is inversely correlated with $L_0$, which means that it increases as $L_0$ decreases; however, in the center of the neutron stars, the opposite trend holds. The cylindrical radial dependence of $\mathcal{B}_{\rm{co}}$ reflects the differential rotation of the core superfluid after a glitch is triggered, as shown in Section~\ref{sec:glitchrise}.
However, in Section \ref{sec:Vela2016}, we also adopt a global $\mathcal{B}_{\rm{co}}$ to examine the overall magnitude of core friction and its role in glitch dynamics.   

\section{Theoretical prediction to glitch rise}\label{sec:glitchrise}

To investigate how the EOS, entrainment effect, and neutron star mass influence glitch rises, we employ the three-component glitch model to simulate the differential rotation of the internal superfluid, crustal frequency evolutions and the corresponding model timing residuals. The observed timing residuals from \citet{Palfreyman2018_Nature556-219}, as reproduced in \citet{Graber2018_ApJ865-23}, are utilized for comparison.

\subsection{Differential rotation of the internal superfluid}\label{subsec:differerntialtotation}

\begin{figure}
\centering
\includegraphics[width=0.45\textwidth]{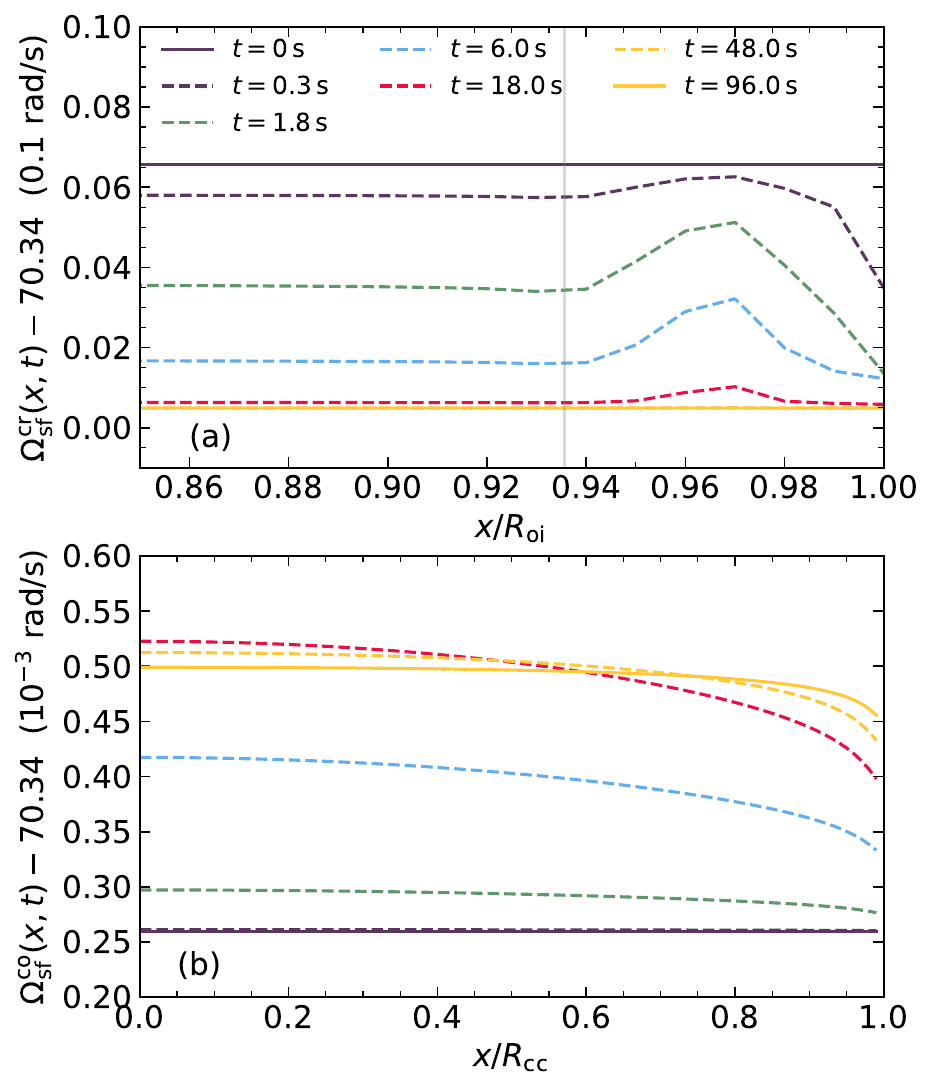}
\caption{(a) Crustal superfluid angular velocity ($\Omega_{\rm{sf}}^{\rm{cr}}$) and (b) core superfluid angular velocity ($\Omega_{\rm{sf}}^{\rm{co}}$) as a function of the normalized cylindrical radius and time following a glitch. The gray solid vertical line mark the radial position of the crust–core interface in the equatorial plane. The effective interaction PKDD with $L_0=40$ MeV, effective mass $m_{\mathrm{n}}^*/m_{\mathrm{n}}=1.0$, and neutron star mass $1.4M_{\odot}$ are adopted.}
\label{fig:Omega}
\end{figure}

In Figure \ref{fig:Omega}, we show the differential rotation evolutions of crust and core superfluids of $1.4M_{\odot}$ neutron star. The effective interaction PKDD with $L_0=40$ MeV is adopted and there is no entrainment effect. In Figure \ref{fig:Omega}(a), the spatial and temporal evolution of $\Omega_{\rm{sf}}^{\rm{cr}}$ are visualized. We observe that the crust superfluid substantially couples with the ``crust" approximately 96 seconds after the glitch, except for the weak coupling region. Smaller values of $\mathcal{B}_{\rm{cr}}$ correspond to slower coupling, particularly in the weak pinning region. The coupling is faster in the region near $x=R_{\rm{oi}}$ due to the largest $\mathcal{B}_{\rm{cr}}$, although this region transfers little angular momentum according to Eq. (\ref{eq:Omegadot_crust}) due to its small superfluid density. 

The core superfluid exhibits even more dramatic dynamics, as shown in Figure \ref{fig:Omega}(b). 
The strong mutual friction (high $\mathcal{B}_{\rm{co}}$) in the central region leads to a prompt core superfluid-``crust'' coupling and rapid angular momentum transfer, manifested as an overshoot where the central superfluid's angular velocity exceeds equilibrium value. To reach global equilibrium, the angular momentum is redistributed to other layers of the core with lower $\mathcal{B}_{\rm{co}}$ on a longer timescale.  The slowest coupling in the region near $x=R_{\rm{cc}}$ (corresponding to smallest $\mathcal{B}_{\rm{co}}$) governing the final stage to equilibrium.
This region and the weak coupling region in the crust require more time to reach equilibrium with the ``crust" component, but their impact on the observable crustal spin frequency is already minimal. We estimate the internal equilibrium timescale $t_{\mathrm{eq}}$ following the trigger of a glitch to be $\sim100$ s, which separates the equilibrium state from the non-equilibrium state. 
Recent three-dimensional simulations of the post-glitch dynamics in the neutron star core indicate that the coupling timescale of the core superfluid depends on the depth within the core \citep{Fuentes2024_ApJ974-300}. A detailed analysis of these effects will be presented in a forthcoming study.

The spatial and temporal evolutions of $\Omega_{\rm{sf}}^{\rm{cr}}$ and $\Omega_{\rm{sf}}^{\rm{co}}$ shown in Figure \ref{fig:Omega} exhibit qualitatively similar patterns across different effective interactions and strengths of entrainment effect. A notable difference is that, as $L_0$ varies, the slowly coupled region in the crust shifts in a manner consistent with the weak coupling region of $\mathcal{B}_{\rm{cr}}$.

\begin{figure}
\centering
\includegraphics[width=0.45\textwidth]{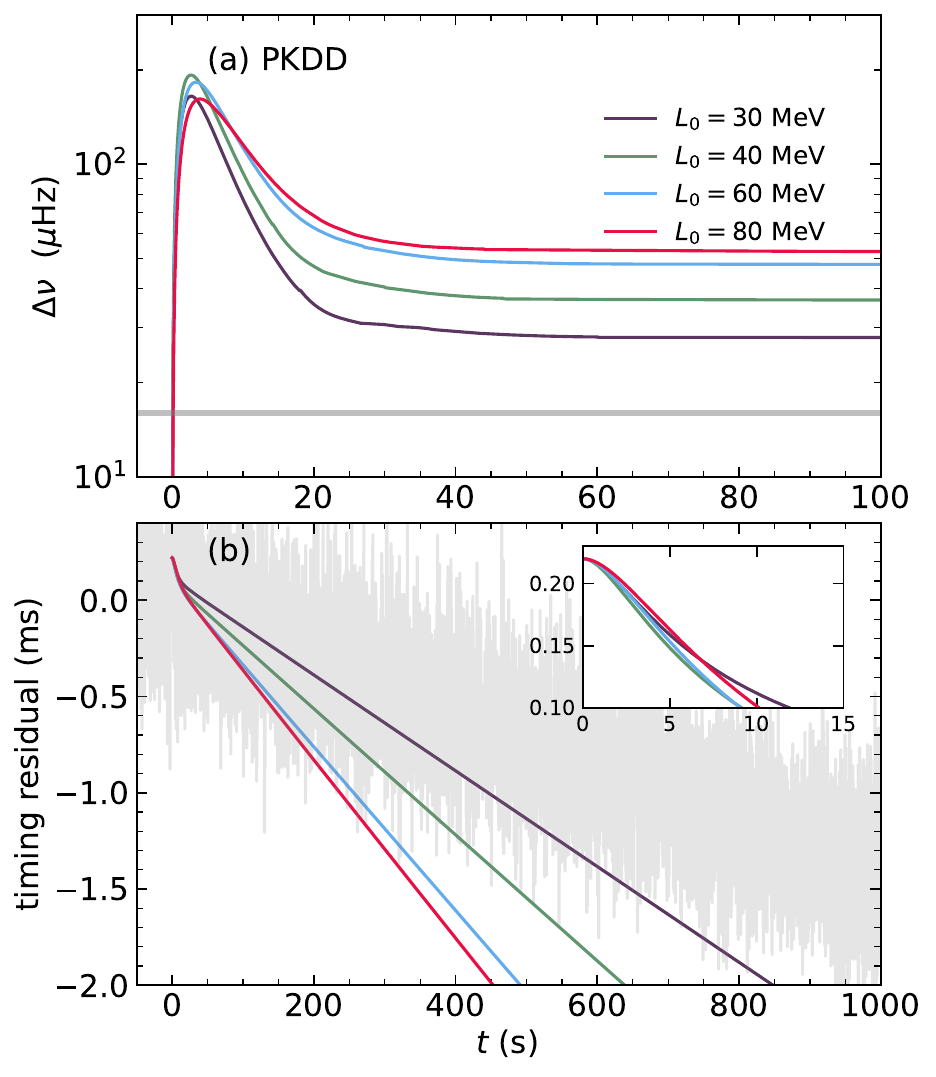}
\caption{
Glitch rise dynamics for different $L_0$ values using the PKDD interaction. (a) Temporal evolution of the crustal frequency shift $\Delta\nu$, showing an overshoot whose peak and timing depend on 
$L_0$. The gray line indicates the final glitch step size 16.02 $\mu$Hz \citep{Palfreyman2018_Nature556-219}. 
(b) Corresponding model timing residuals compared to observations from \citet{Palfreyman2018_Nature556-219} and \citet{Graber2018_ApJ865-23}. The inset highlights the first 15 s, demonstrating the model's sensitivity to $L_0$ during the initial rise.
}
\label{fig:Delta_nu_L_PKDD}
\end{figure}

\subsection{Glitch rise and crustal frequency evolution}\label{subsec:crustalfrequency}

In Figure \ref{fig:Delta_nu_L_PKDD}(a), we plot the evolution of crustal frequency $\Delta\nu = [\Omega_{\rm{p}}(t)-\Omega_{\rm{p}}(0)]/2\pi$ for a $1.4M_{\odot}$ neutron star calculated using PKDD with different $L_0$ in the absence of entrainment effect. We can clearly see that an overshoot appears just after the glitch. $\Delta\nu$ rapidly increases to a peak value and then decays to an equilibrium value $\Delta\nu_{\text{eq}}$. The peak value of crustal frequency is determined by the relative strength between the crust and core mutual friction. 
For all our present simulations, the glitch rise time are significantly less than the observed upper limit of 12.6 s. 

The equilibrium value $\Delta\nu_{\rm{eq}}$ can be determined by the conservation of angular momentum. Neglecting the external torque, for any two times $t_1<t_2$, the equations (\ref{eq:Omegadot_sf})--(\ref{eq:Omegadot_crust}) yield:
\begin{align}
\Omega_{\rm{p}}(t_2) - \Omega_{\rm{p}}(t_1) 
&= \frac{2\pi}{I} \Bigg[
 \int dx \, x^3 b_{\rm cr}(x) \left( \omega_{\rm cr}(x,t_1) - \omega_{\rm cr}(x,t_2) \right) \notag\\
& + \int dx \, x^3 b_{\rm co}(x) \left( \omega_{\rm co}(x,t_1) - \omega_{\rm co}(x,t_2) \right) \Bigg], \label{eq:Omega_p_t2_t1_integ}
\end{align}
where $\omega=\Omega_{\rm{sf}}-\Omega_{\rm{p}}$ is the lag between superfluid and ``crust'' component. For the initial state at $t_1=0$, our initial conditions have assumed that the lags $\omega_{\rm{cr}}$ and $\omega_{\rm{co}}$ are spatially uniform. For the equilibrium state at $t>t_{\rm{eq}}$, we see from Figure \ref{fig:Omega} that both $\Omega_{\rm{sf}}^{\rm{cr}}$ and $\Omega_{\rm{sf}}^{\rm{co}}$ are approximately spatially uniform at this moment. Therefore, for $t_1 = 0$ and $t_2 = t_{\rm{eq}}$, the Eq. (\ref{eq:Omega_p_t2_t1_integ}) reduce to a scalar form:
\begin{align}
\Omega_{\rm{p}}(t_{\rm{eq}}) - \Omega_{\rm{p}}(0) 
\approx \frac{1}{I} \Bigg[
& I_{\rm{sf}}^{\rm{cr}} \left( \omega_{\rm{cr}}(0) - \omega_{\rm{cr}}(t_{\rm{eq}}) \right) \notag\\
& + I_{\rm{sf}}^{\rm{co}} \left( \omega_{\rm{co}}(0) - \omega_{\rm{co}}(t_{\rm{eq}}) \right) \Bigg]. \label{eq:Omega_p_t2_t1_scalar}
\end{align}
At the initial state, $\omega_{\rm{co}}(0)=0$; at the equilibrium state, $\omega_{\rm{cr}}(t_{\rm{eq}})\approx\omega_{\rm{co}}(t_{\rm{eq}})\approx0$. Eq. (\ref{eq:Omega_p_t2_t1_scalar}) can be further reduced to
\begin{equation}\label{eq:nu_eq}
\Delta\nu_{\mathrm{eq}}\approx\frac{I_{\rm{sf}}^{\rm{cr}}}{I}\frac{\omega_{\rm{cr}}(0)}{2\pi}.
\end{equation}
From equation (\ref{eq:nu_eq}), with a given $\omega_{\rm{cr}}(0)$, the equilibrium value $\Delta\nu_{\text{eq}}$ is approximately determined by the capacity $I_{\rm{sf}}^{\rm{cr}}/I$ for angular momentum transfer during a glitch, and hence depends on EOS, entrainment and neutron star mass.

The crustal frequency evolution $\Delta\nu$ can be integrated to the model timing residual via $r(t) = -2\pi\phi/\Omega_{\rm{p}}(0)$ with the phase $\phi=\int_0^t\Delta\nu\mathrm{d}t'$. The model timing residual is sensitive to the microscopic inputs and neutron star mass, e.g., Figure \ref{fig:Delta_nu_L_PKDD}(b) shows its dependence on $L_0$. Following \citet{Graber2018_ApJ865-23}, the model timing residuals are shifted $\Delta t=0.22$ s at $t=0$ to consider the frequency slowdown preceding the glitch.

\begin{figure}
\centering
\includegraphics[width=0.45\textwidth]{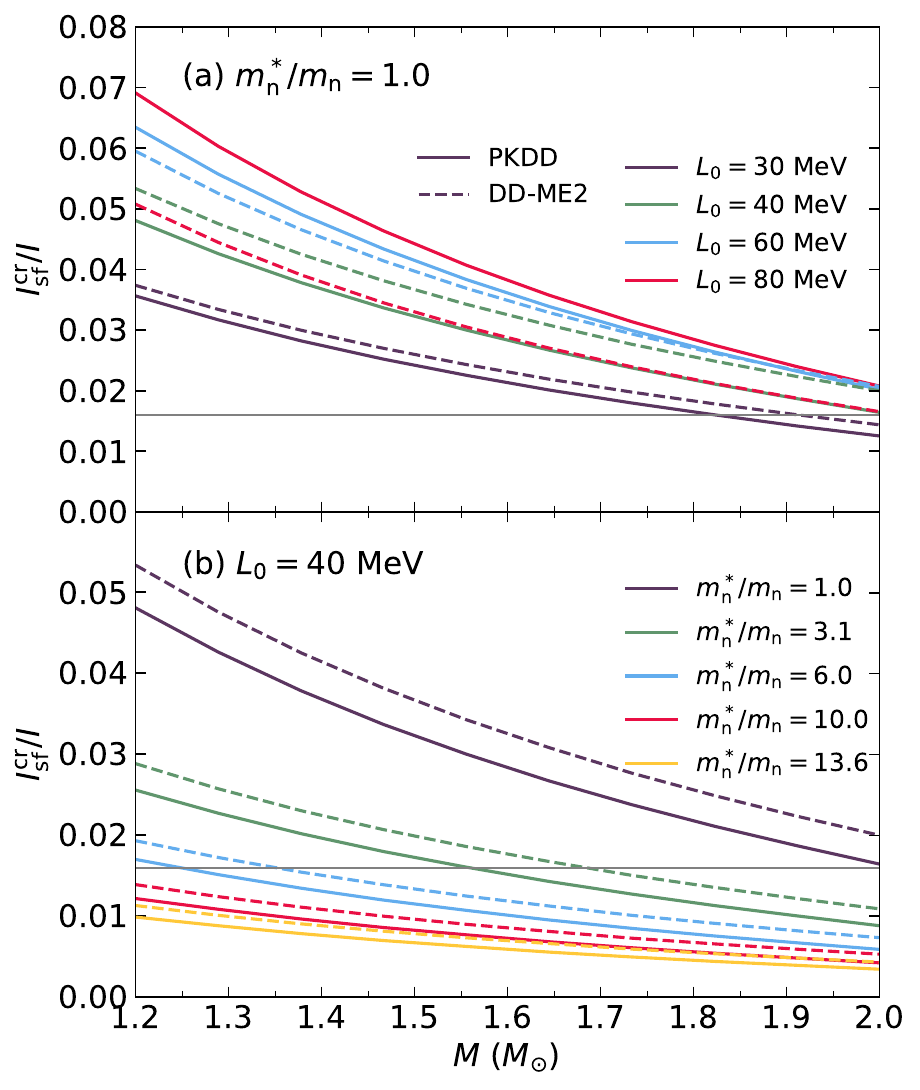}
\caption{
Fraction of the total moment of inertia carried by the crustal superfluid reservoir ($I_{\rm{sf}}^{\rm{cr}}/I$) as a function of neutron star mass. (a) Dependence on the symmetry energy slope $L_0$ for both PKDD (solid lines) and DD-ME2 (dashed lines) interactions. The gray horizontal line marks the value required to explain the glitch step size of the 2016 Vela event. 
(b) Dependence on the effective mass $m_{\mathrm{n}}^*/m_{\mathrm{n}}$, showing how stronger entrainment (larger $m_{\mathrm{n}}^*/m_{\mathrm{n}}$) systematically reduces the available angular momentum reservoir. These trends directly determine the equilibrium glitch step size $\Delta\nu_{\mathrm{eq}}$.}
\label{fig:MoI}
\end{figure}

\begin{figure}
\centering
\includegraphics[width=0.45\textwidth]{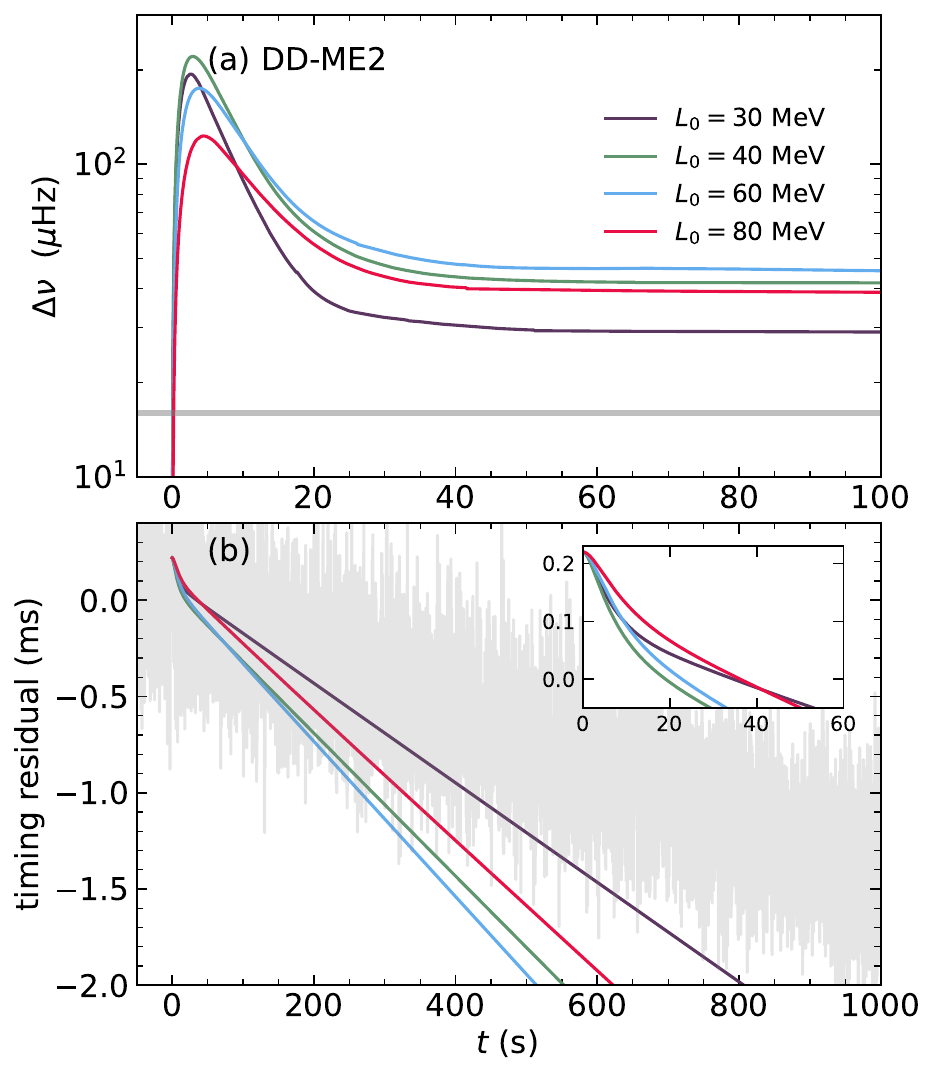}
\caption{
Non-monotonic dependence of glitch dynamics on the symmetry energy slope $L_0$ for the DD-ME2 interaction. (a) Temporal evolution of the crustal frequency shift $\Delta\nu$ shows complex behavior that varies with $L_0$, in contrast to the monotonic trend seen with PKDD (Fig. \ref{fig:Delta_nu_L_PKDD}). (b) Corresponding model timing residuals compared to observations. The inset highlights the first 60 s, where the non-monotonic dependence creates distinctive residual patterns. This contrasting behavior between nuclear interactions provides a potential discriminant for constraining the nuclear EOS.}
\label{fig:Delta_nu_L_DD-ME2}
\end{figure}

\subsection{Effects of EOS, entrainment, and neutron star mass on the glitch rise}\label{subsec:effects}

Here we discuss how $L_0$ affects the crustal frequency evolutions and the model timing residuals. In the non-equilibrium state, as we can see in Figure \ref{fig:Delta_nu_L_PKDD} (a), as $L_0$ increases from 40 MeV to 80 MeV, the peak value in the overshoot decreases and the location of the peak moves right. On the one hand, larger $L_0$ predicts a smaller $\mathcal{B}_{\rm{cr}}$ near $x=R_{\rm{cc}}$ and at $x<R_{\rm{cc}}$, resulting in slower rise and smaller peak value; on the other hand, larger $L_0$ also predicts a larger $\mathcal{B}_{\rm{co}}$ in the region except near $x=R_{\rm{cc}}$, leading to faster decay and smaller peak value. Consequently, as $L_0$ increases from 40 MeV to 80 MeV, the peak value of the overshoot decreases and its corresponding time lengthens. PKDD with $L_0=30$ MeV predicts non-monotonic $\mathcal{B}_{\rm{cr}}$ near $x=R_{\rm{cc}}$, causing the shape of $\Delta\nu$ to deviate from the pattern we described for $L_0>40$ MeV. The complex shape of the crustal frequency in the overshoot that causes the dependence $L_0$ of the model timing residuals is not trivial during the initial dozen seconds, as we can see in the inset of Figure \ref{fig:Delta_nu_L_PKDD} (b). 
 
Figure \ref{fig:MoI} shows the model-predicted dependence of the superfluid angular momentum reservoir on the neutron star mass, the symmetry energy slope $L_0$, and the entrainment strength.
Panel (a) shows a critical divergence between the two EOS families: $I_{\rm{sf}}^{\rm{cr}}/I$ increases monotonically with $L_0$ for PKDD, while it peaks at intermediate $L_0$ (40-80 MeV) for DD-ME2. This non-monotonic dependence of the DD-ME2 interaction also leads to a non-monotonic dependence of $\Delta\nu_{\rm{eq}}$ on $L_0$, as shown in Figure \ref{fig:Delta_nu_L_DD-ME2}. Figure \ref{fig:MoI}(b) shows that stronger entrainment systematically reduces $I_{\rm{sf}}^{\rm{cr}}/I$ for all neutron star masses considered. This reduction directly decreases the available angular momentum reservoir and therefore lowers the predicted equilibrium glitch amplitude $\Delta\nu_{\mathrm{eq}}$.

\begin{figure}
\centering
\includegraphics[width=0.45\textwidth]{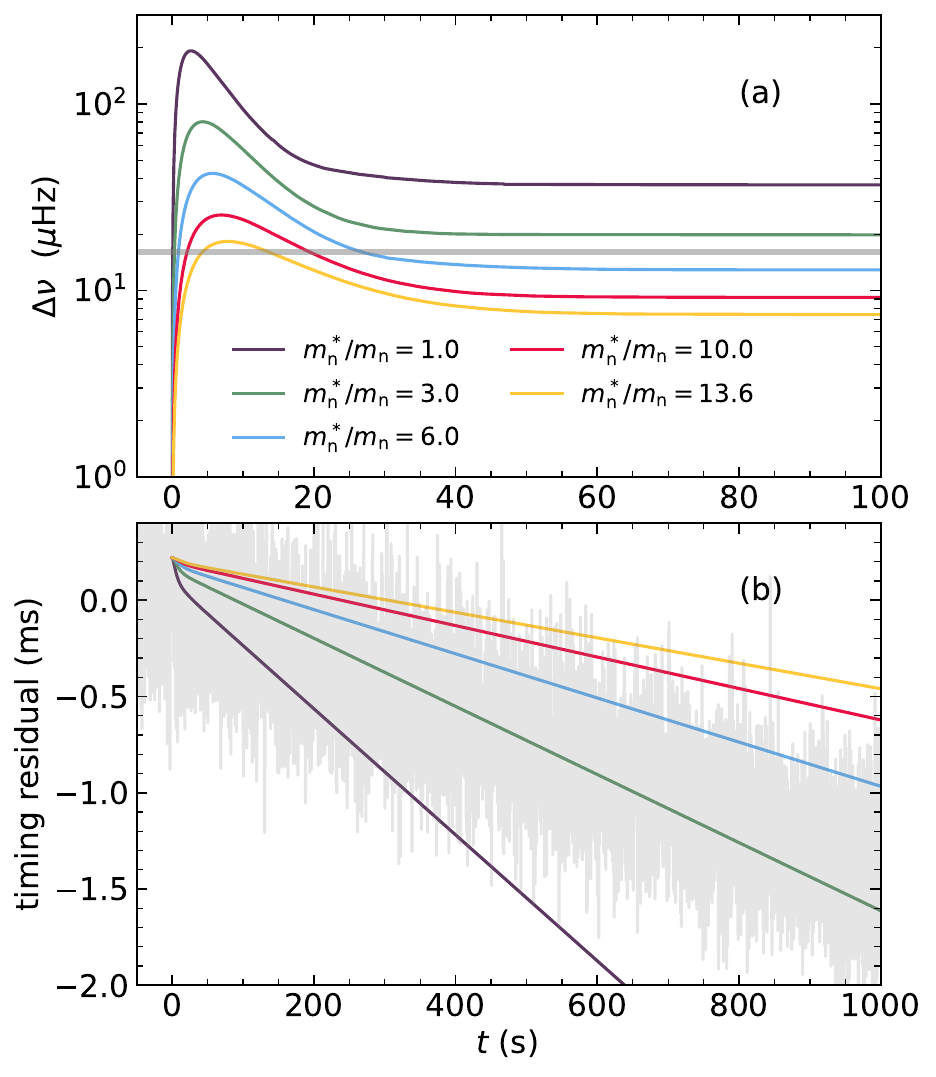}
\caption{Systematic suppression of glitch dynamics by entrainment effects. (a) Crustal frequency shift $\Delta\nu$ shows reduced overshoot amplitude and smaller equilibrium values with stronger entrainment (larger $m_{\mathrm{n}}^*/m_{\mathrm{n}}$). (b) Model timing residuals demonstrate high sensitivity to the entrainment strength $m_{\mathrm{n}}^*/m_{\mathrm{n}}$.}
\label{fig:Delta_nu_fi}
\end{figure}

In Figure \ref{fig:Delta_nu_fi}(a), we demonstrate how the entrainment effect affects the shape of $\Delta\nu$ and the model timing residuals. In the non-equilibrium state, the peak of the overshoot is systematically suppressed because $\mathcal{B}_{\rm{cr}}$ is weakened by stronger entrainment effect, see Figure \ref{fig:friction_crust}(b), while $\mathcal{B}_{\rm{co}}$ remains unchanged. Consequently, the rise timescale lengthens, whereas the decay rate stays constant. In the equilibrium state, $\Delta\nu_{\mathrm{eq}}$ decreases with enhanced entrainment, which can be simply understood as entrainment reducing the capacity of the angular momentum reservoir (i.e., $I_{\rm{sf}}^{\rm{cr}}/I$, see Figure \ref{fig:MoI}(b)), thereby decreasing angular momentum transfer and resulting in smaller glitch step sizes. As depicted in Figure \ref{fig:Delta_nu_fi}(b), the model timing residuals are highly sensitive to the entrainment effect, providing a potential method to constrain the strength of the entrainment effects in neutron star interiors.

\begin{figure}
\centering
\includegraphics[width=0.45\textwidth]{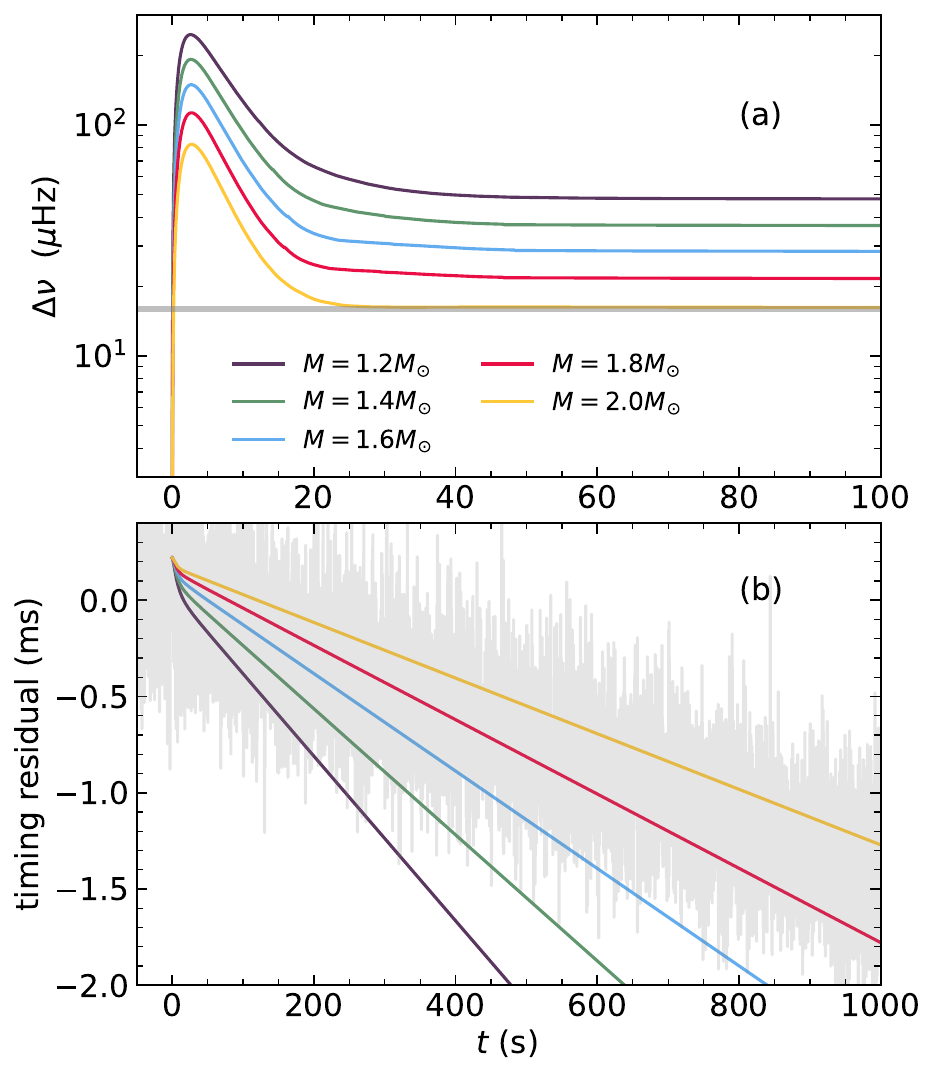}
\caption{NS mass as a determining factor in glitch characteristics. (a) Higher mass neutron stars exhibit suppressed overshoot and smaller glitch steps. (b) The corresponding timing residuals show distinct mass-dependent signatures, particularly in the overshoot region.}
\label{fig:Delta_nu_M}
\end{figure}

In Figure \ref{fig:Delta_nu_M} (a), we show how the neutron star mass affects the shape of $\Delta\nu$ and the model timing residuals. In the non-equilibrium state, the peak of the overshoot is systematically suppressed because the core region with strong mutual friction expands as the neutron star mass increases. In the equilibrium state, $\Delta\nu_{\mathrm{eq}}$ decreases with neutron star mass, which can be simply understood as the capacity of the angular momentum reservoir is reduced due to the thinner inner crust, thus decreasing angular momentum transfer and resulting in smaller glitch step sizes. As depicted in Figure \ref{fig:Delta_nu_M}(b), the model timing residuals are highly sensitive to the neutron star mass, providing a potential method to measure the neutron star mass by the timing observation of the pulsar.

Note that in the present study, we assume that the angular momentum reservoir is fully depleted after triggering a glitch. The uncertainty associated with the fraction of unpinned vortices is not considered here. Realistically modeling this fraction remains an open challenge in glitch theory~\citep[e.g.,][]{Antonelli2020_MNRAS499-3690}.

\section{Model fitting to the 2016 Vela glitch}\label{sec:Vela2016}

We have confirmed that the EOSs, entrainment effect, and neutron star mass significantly affect the glitch rises. In this section, within the adopted cylindrical geometry mapping, we try to use the observed timing residuals and our three-component model to explore whether the glitch rise can encode information on microscopic parameters. We emphasize that our results do not constitute a robust astrophysical inference, but rather serve as a controlled experiment within a restricted model framework.

\subsection{MCMC simulations and parameter correlations}\label{subsec:mcmc}

\begin{figure*}
\centering
\includegraphics[width=1.0\textwidth]{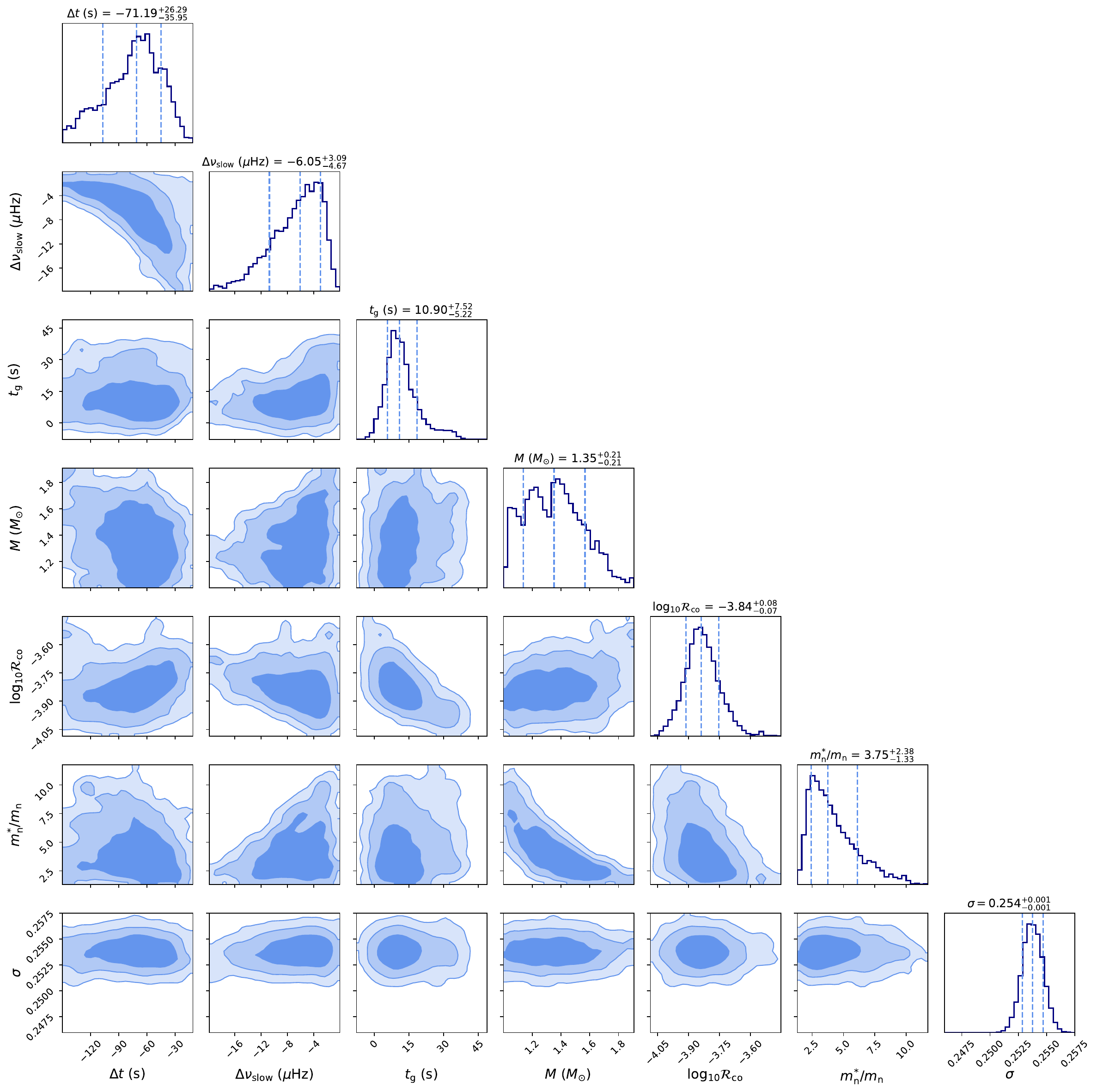}
\caption{Posterior distributions and correlations of model parameters from the MCMC fit to the 2016 Vela glitch data, using the DD-ME2 interaction with $L_0=60$ MeV. The anti-correlation between the pre-glitch slowdown duration $\Delta t$ and magnitude $\Delta\nu_{\mathrm{slow}}$, and the anti-correlation between Vela mass $M$ and entrainment strength, are clearly visible.
}
\label{fig:corner_PKDD_L40}
\end{figure*}

To determine the parameters of the three-component glitch model by fitting the 2016 Vela glitch, we first extended the model to incorporate the observed pre-glitch crustal slowdown, denoted as $\Delta\nu_{-}$.
This transient slowdown ($\Delta\nu_{-} = 5.40^{+3.39}_{-2.05}\ \mu$Hz over approximately 100 s; \citet{Ashton2019_NA36-844}) could represent a stochastic fluctuation to trigger the glitch \citep{Ashton2019_NA36-844} or a signal of the formation of a new vortex trap zone by a crust-breaking quake \citep{Guegercinoglu2020_MNRAS496-2506}.
In this work, we follow the treatment in \citet{Ashton2019_NA36-844} by phenomenologically modeling the slowdown in the crustal frequency using a rectangular function, where the triggering of the glitch immediately follows the slowdown event. The complete model is therefore expressed as: 
\begin{equation}
\begin{aligned}\label{eq:fullmodel}
    \Delta\nu(t) = & \Delta\nu_{\mathrm{slow}}\Pi(t;t_{\mathrm{g}}+\Delta t, t_{\mathrm{g}}^{-}) \\ &+ (\Delta\nu_{3c}-\Delta\nu_{\mathrm{slow}})H(t-t_{\mathrm{g}}),
\end{aligned}
\end{equation}
where $t_{\mathrm{g}}$ is the glitch time, $\Pi$ is the rectangle function such 
that the frequency decreases by $\Delta\nu_{\mathrm{slow}}$ for the period $|\Delta t|$ prior the glitch. Note that in our simulations, the crustal frequency does not recover to its pre-slowdown value. $\Delta\nu_{\mathrm{3c}}$ represents the post-glitch frequency evolution calculated through the three-component model. $H(t-t_{\mathrm{g}})$ is the Heaviside step function. In the three-component glitch model, we treat $\mathcal{R}_{\rm{co}}$ as a free parameter to examine the overall magnitude of core friction and its role in glitch dynamics. Our fittings are performed across all effective interactions presented in Section \ref{subsec:EoS}. There are six model parameters to be determined in our model: $\Delta t$, $\Delta\nu_{\mathrm{slow}}$, $t_{\mathrm{g}}$, $M$, $\mathcal{R}_{\rm{co}}$, and $m_{\mathrm{n}}^*/m_{\mathrm{n}}$.

We investigated the rise phase of the 2016 Vela glitch by implementing Eq.~(\ref{eq:fullmodel}) within an MCMC framework, employing the \texttt{emcee} ensemble sampler to perform parameter estimation.
The model $\Delta\nu(t)$ is integrated to generate the model timing residuals $r(t)$, which are subsequently fitted against the observed timing residuals. The priors of $\Delta t$, $\Delta\nu_{\mathrm{slow}}$, and $t_{\mathrm{g}}$ are selected as uniform distributions based on the estimation of the previous work \citep{Ashton2019_NA36-844}: $\Delta t(\mathrm{s})\sim \mathrm{Uniform}[-100,-10]$, $\Delta\nu_{\mathrm{slow}}(\mu\mathrm{Hz})\sim \mathrm{Uniform}[-10,0]$, and $t_{\mathrm{g}}(\mathrm{s})\sim \mathrm{Uniform}[-10,50]$. The prior of neutron star mass is the uniform distribution based on the observed upper limit \citep{Romani2022_ApJL934-L17}: $M(M_{\odot})\sim\mathrm{Uniform}[1.0,2.3]$. 
For the prior of $\mathcal{B}_{\rm{co}}$, we adopt a uniform prior of the logarithmic drag parameter $\log_{10}\mathcal{R}_{\rm{co}}$ : $\log_{10}\mathcal{R}_{\rm{co}}\sim\mathrm{Uniform}[-5.0,-3.0]$. For the strength of the entrainment effect, we adopt a uniform prior distribution $m_{\mathrm{n}}^*/m_{\mathrm{n}}\sim\mathrm{Uniform}[1.0,13.6]$, where the upper limit is the theoretical value at about $0.03~\rm{fm}^{-3}$ from \citet{Chamel2012_PRC85-035801}, and the lower limit corresponds to the case without the entrainment effect. 
An additional parameter is the uncertainty $\sigma$ on single measurement of the time residual. 
Assuming that the influence of $\sigma$ on the correlation between timing residuals and pulse arrival times is negligible, we adopt a uniform prior $\sigma \sim \mathrm{Uniform}[0.2, 0.3]$. Under the assumption that individual pulse measurements are statistically independent, the likelihood function can be expressed as \citet{Ashton2019_NA36-844} and \citet{Montoli2020_A&A642-A223} as:
\begin{equation}
\begin{aligned}\label{eq:likelihood}
\mathcal{P}\sim\prod_i\frac{1}{\sqrt{2\pi\sigma^2}}e^{-\frac{(r(t_i)-r_i)^2}{2\sigma^2}},
\end{aligned}
\end{equation}
where $r_i$ and $t_i$ are the single measured timing residual and corresponding time. 

We present the posterior distributions of the model parameters and their uncertainty in Figure \ref{fig:corner_PKDD_L40}. Within the adopted three-component glitch model and priors, statistically significant correlations are inferred for the parameter pairs ($\Delta t$, $\Delta\nu_{\mathrm{slow}}$) and ($M$, $m_{\mathrm{n}}^*/m_{\mathrm{n}}$).
The inferred anti-correlation between $\Delta t$ and $\Delta\nu_{\mathrm{slow}}$ is consistent with the results reported by \citet{Ashton2019_NA36-844}. These two parameters are introduced to model the elevated timing residuals caused by the slowdown prior to the glitch, and both $\Delta t$ and $\Delta\nu_{\mathrm{slow}}$ contribute positively to the amplitude of the timing residual. Consequently, an increase in one parameter necessarily requires a decrease in the other. 
Within the framework of our glitch model, the inferred anti-correlation between $M$ and $m_{\mathrm{n}}^*/m_{\mathrm{n}}$ implies that, for a fixed observed glitch step size, a stronger entrainment effect (larger $m_{\mathrm{n}}^*/m_{\mathrm{n}}$) requires a lower stellar mass to provide a sufficient angular momentum reservoir, see Figure \ref{fig:MoI}.
 
\begin{table*}
  \centering
  \caption{Marginalized posterior of model parameters fitted to the 2016 Vela glitch for DD-ME2 and PKDD with varying values of $L_0$. The results are given by the median values with 16\%--84\% credible intervals.}\label{tab:fittings}
  \setlength\tabcolsep{4pt}
  \begin{tabular}{lrrrrrrrr}
    \hline
    \hline
     & \multicolumn{4}{c}{DD-ME2} & \multicolumn{4}{c}{PKDD} \\
     \cmidrule(r){2-5} \cmidrule(r){6-9}
     $L_0$ (MeV) & 30 & 40 & 60 & 80 & 30 & 40 & 60 & 80 \\
    \hline
    $\Delta t$ (s) & $-80.56_{-33.73}^{+29.51}$ & $-81.93_{-35.35}^{+28.80}$ & $-71.19_{-35.95}^{+26.29}$ & $-87.34_{-33.37}^{+27.86}$ & $-81.92_{-38.65}^{+27.73}$ & $-78.01_{-38.76}^{+28.31}$ & $-69.25_{-34.54}^{+25.75}$ & $-69.50_{-39.91}^{+25.21}$ \\
    $\Delta\nu_{\mathrm{slow}}$ ($\mu$Hz) & $-4.22_{-2.68}^{+2.02}$ & $-4.31_{-2.66}^{+1.95}$ & $-6.05_{-4.67}^{+3.09}$ & $-4.30_{-2.88}^{+2.01}$ & $-4.14_{-3.64}^{+2.25}$ & $-4.78_{-3.94}^{+2.38}$ & $-6.47_{-4.94}^{+3.16}$ & $-6.03_{-4.43}^{+3.03}$ \\
    $t_{\mathrm{g}}$ (s) & $16.41_{-7.09}^{+18.53}$ & $15.17_{-6.42}^{+16.94}$ & $10.90_{-5.22}^{+7.52}$ & $13.62_{-6.59}^{+11.90}$ & $14.66_{-6.24}^{+19.56}$ & $14.88_{-6.20}^{+18.72}$ & $11.91_{-5.01}^{+9.76}$ & $11.42_{-5.09}^{+9.02}$ \\
    $M$ ($M_{\odot}$) & $1.15_{-0.09}^{+0.19}$ & $1.37_{-0.22}^{+0.21}$ & $1.35_{-0.21}^{+0.21}$ & $1.34_{-0.18}^{+0.19}$ & $1.16_{-0.12}^{+0.20}$ & $1.33_{-0.21}^{+0.16}$ & $1.44_{-0.21}^{+0.20}$ & $1.51_{-0.18}^{+0.16}$ \\
    $\log_{10}\mathcal{R}_{\rm{co}}$ & $-3.84_{-0.10}^{+0.11}$ & $-3.84_{-0.09}^{+0.09}$ & $-3.84_{-0.07}^{+0.08}$ & $-3.89_{-0.07}^{+0.08}$ & $-3.81_{-0.11}^{+0.11}$ & $-3.83_{-0.10}^{+0.10}$ & $-3.83_{-0.08}^{+0.08}$ & $-3.85_{-0.07}^{+0.08}$ \\
    $m_{\mathrm{n}}^*/m_{\mathrm{n}}$ & $3.15_{-1.07}^{+0.94}$ & $3.65_{-1.12}^{+2.31}$ & $3.75_{-1.33}^{+2.38}$ & $3.25_{-1.03}^{+1.74}$ & $2.74_{-0.98}^{+1.02}$ & $3.20_{-0.97}^{+1.54}$ & $3.37_{-1.29}^{+1.92}$ & $3.28_{-1.15}^{+1.93}$ \\
    \hline     \hline
  \end{tabular}
\end{table*}

\subsection{Constraints on the Vela's mass, EOS, and interiors}\label{subsec:constrain to vela}

Table \ref{tab:fittings} summarizes the marginalized posterior distributions of the model parameters for each effective interaction and geometry mapping considered. 
Overall, $\Delta t$ exhibits a decreasing trend with $L_0$, while $\Delta\nu_{\mathrm{slow}}$ shows a corresponding increasing trend due to their intrinsic correlation. The trigger time of the glitch exhibits an anti-correlation with $L_0$, indicating that the inferred $t_{\mathrm{g}}$ from observations can serve as an additional probe for constraining the EOS. 

\begin{figure}
\centering
\includegraphics[width=0.45\textwidth]{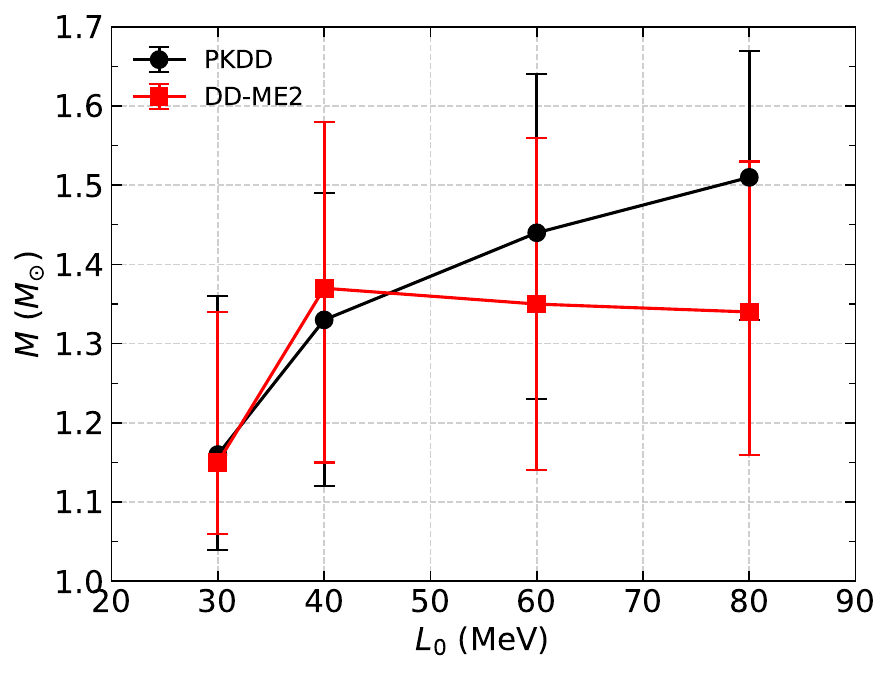}
\caption{The Vela masses with 16\%--84\% credible intervals, inferred from MCMC simulations, as a function of $L_0$ for DD-ME2 and PKDD.
}
\label{fig:postM}
\end{figure}

The Vela mass exhibits a consistent $L_0$-dependencies under two set of effective interactions, which is visualized in Figure \ref{fig:postM}. For PKDD, the inferred Vela masses increases monotonically with increasing $L_0$; for DD-ME2, the dependence of inferred Vela mass on $L_0$ shows a non-monotonic behavior, first increasing and then decreasing, which is consistent with the trend of $I_{\rm{sf}}^{\rm{cr}}/I$ as a function of $L_0$. Under our restricted modeling framework, the mass of the Vela pulsar is estimated as 1.05–1.70 $M_{\odot}$. This is broadly consistent with the constraint of 1.25–1.45 $M_{\odot}$ from the maximum glitch size \citep{Pizzochero2017_NA1-0134} and is close to the lower limit inferred from the 2000 glitch by \citet{Shang2021_ApJ923-108}. However, it is in tension with the high-mass solution (>2.0 $M_{\odot}$) found by \citet{Tu2025_ApJ984-200} for the 2000 glitch using the snowplow model. 

The entrainment effect first strengthens and then weakens with increasing $L_0$ both for DD-ME2 and PKDD. The weak entrainment effect ($m_{\mathrm{n}}^*/m_{\mathrm{n}}\sim1.76$--$6.13$) inferred from our present study supports the theoretical calculation in \citet{2025PhRvL.135m2701A}.

For both DD-ME2 and PKDD, the core mutual friction coefficients $\mathcal{B}_{\rm{co}}$ are weakly sensitive to the EOS.
The inferred $\mathcal{B}_{\rm{co}}$ is on the order of $10^{-4}$ ($\log_{10}\mathcal{R}_{\rm{co}}\sim-3.8$), which is consistent in magnitude with that given by Eq.~(\ref{eq:friction_core}), see Figure \ref{fig:friction_core}.
Noted that Eq.~(\ref{eq:friction_core}) describing $\mathcal{B}_{\rm{co}}$ is applicable to the outer core region, where the density is typically $\rho \sim 10^{14}~\mathrm{g\,cm^{-3}}$ and the proton fraction is $Y_{\mathrm{p}} \sim 0.05$.
Since the Vela mass inferred from our fitting is relatively low, the corresponding core densities do not deviate significantly from this validity range. 

Using a three-component glitch model with phenomenological coupling parameters and fractional MoIs, \citet{Montoli2020_A&A642-A223} derived posterior distributions for the effective crustal and core mutual friction coefficients, $\langle\mathcal{B}_{\mathrm{crust}}\rangle$ and $\langle\mathcal{B}_{\mathrm{core}}\rangle$. Our approach instead parameterize the inference at the microphysical level, from which $\mathcal{B}_{\rm{cr}}$ is calculated self-consistently, yielding $\sim10^{-3}$--$10^{-2}$. Our posterior gives $1-\epsilon_{\rm{n}}\approx 3$ for the inner crust. Under this condition, our crustal mutual friction coefficients agree, to order of magnitude, with $\langle\mathcal{B}_{\mathrm{crust}}\rangle \approx 2.4 \times 10^{-3}$--$0.7$ reported by \citet{Montoli2020_A&A642-A223} for $\langle1-\epsilon_{\rm{n}}\rangle_{\rm{crust}}\approx 4$. For the core, our $\mathcal{B}_{\rm{co}}$ is about an order of magnitude larger than the effective value in \citet{Montoli2020_A&A642-A223}. However, since they did not directly sample $\langle\mathcal{B}_{\mathrm{core}}\rangle$ but constrained a phenomenological coupling parameter $b_2$ related to it via an assumed entrainment coefficient, a direct quantitative comparison is not straightforward. Furthermore, core entrainment is not included in our present framework. Incorporating core entrainment in a self-consistent manner, therefore, remains an important extension for future studies.

\section{Summary and perspective}\label{sec:summary}

In this study, we have examined the superfluid coupling dynamics during the 2016 Vela glitch within a three-component glitch framework, using unified EOSs to compute the internal structure and microphysical inputs of neutron stars. Our aim has been exploratory: to assess whether, and in what way, the morphology of the glitch rise could, in principle, encode information about microphysical parameters such as mutual friction and entrainment, under a set of restrictive modeling assumptions.

For this purpose, we refitted DD-ME2 and PKDD to nuclear saturation properties and applied them to construct several unified EOSs, enabling self-consistent model calculations of the interior composition, superfluidity, and global properties of neutron stars. 
On this basis, we calculated crustal mutual friction coefficients $\mathcal{B}_{\rm{cr}}$ from Kelvin-wave excitation between vortices and nuclear clusters, and core mutual friction coefficients $\mathcal{B}_{\rm{co}}$ assuming dissipation dominated by electron scattering off magnetized vortices. We then embedded these coefficients in a simplified three-component model, extended to include differential rotation of the core superfluid, and explored how three key input, i.e., the symmetry energy slope $L_0$, an effective entrainment parameter $m_{\mathrm{n}}^*/m_{\mathrm{n}}$, and the stellar mass, affect the glitch rise and associated timing residuals.

Within this framework, the density dependence of the mutual friction coefficient leads to complex coupling dynamics within the neutron stars. 
Specifically, the stronger mutual friction coefficient in the central region of the neutron stars produces an overshoot feature there. 
The overshoot in the crustal frequency is jointly determined by both the crustal and core mutual friction coefficients. Except for the case of $L_0=30$ MeV, increasing $L_0$ systematically reduces the crustal mutual friction coefficient near $x=R_{\rm{cc}}$ and at $x<R_{\rm{cc}}$ but enhances it in deeper core regions, resulting in a decrease in the peak of the overshoot in the modeled glitch evolution. 
The $L_0$-dependence of the capacity of the angular momentum reservoir depends on the EOSs, resulting in two distinct correlations between the predicted glitch step size and $L_0$: monotonic increase (PKDD) and increase-then-decrease (DD-ME2). These behaviors reflect differences in the isoscalar channel of the underlying effective interactions. A stronger entrainment effect reduces both the crustal mutual friction coefficient and the capacity of the modeled angular momentum reservoir capacity, causing both a lower peak of the overshoot and a smaller glitch step size. 
Similarly, a larger neutron star mass corresponds to reduced capacity of the angular momentum reservoir in the model, resulting in a lower peak of the overshoot and a smaller predicted glitch step size. 

We performed MCMC simulations of the timing residuals of the 2016 Vela glitch using this specific three-component glitch model. Within this modeling framework, the vela mass is estimated as 1.05--1.70 $M_{\odot}$; the core mutual friction coefficient, $\mathcal{B}_{\rm{co}}$, is on the order of $10^{-4}$; the inner-crust entrainment parameter $1-\epsilon_{\rm{n}}$ favor relatively weak entrainment. Although these results do not constitute robust observational inferences, they clearly demonstrate that the glitch rise can, in principle, carry information about both the microphysics and the stellar mass, thereby laying the groundwork for model-independent observational inferences in future work.

Our analysis indicates, within this specific three-component glitch model and the adopted unified EOSs, an anti-correlation between the inferred Vela mass and the strength of the entrainment effect.
The dependence of Vela mass on $L_0$ exhibits two distinct behaviors: monotonic increase for PKDD and an increase followed by a decrease for DD-ME2. 
Previously, using the same effective interactions and self-consistently calculated crustal pinning forces, \citet{Tu2025_ApJ984-200} applied the snowplow model to the 2000 Vela glitch and inferred a Vela mass exceeding 2.0$M_{\odot}$.  
Two different glitch models yield divergent estimates of the Vela mass. The difference between the two studies can be attributed to several methodological and physical assumptions:
(1) \citet{Tu2025_ApJ984-200} did not include the entrainment effect, which can substantially reduce the effective superfluid MoI;
(2) the snowplow model considers only the unpinning of a fraction of vortices and a limited portion of the inner crust’s MoI, while our treatment includes a more complete coupling of components;
(3) the two glitches may have been triggered by distinct physical mechanisms—for instance, the 2016 Vela glitch analyzed here may have been initiated by a crustal fracture (starquake); and
(4) \citet{Tu2025_ApJ984-200} constrained the model parameters by fitting the step changes in spin frequency and its derivative about one minute after the glitch, whereas our analysis focuses on the detailed timing residuals, capturing the early relaxation dynamics.

Our present results should not be interpreted as delivering quantitative constraints on the neutron star EOS, mutual friction mechanisms, or entrainment strength. They rest on strong idealizations (e.g., axisymmetry, cylindrical mapping of a 3D star, straight rigid vortices, prescribed forms of pinning and friction, neglect of core entrainment, and the assumption of full reservoir depletion at each glitch) that are unlikely to be fully realized in nature. Instead, the present analysis is best regarded as a controlled experiment: it shows that, within a particular simplified dynamical setting, the observed glitch rise of the 2016 Vela glitch are sensitive to specific combinations of microphysical parameters, and that degeneracies and correlations among these parameters naturally emerge.

Looking forward, a key implication of this work is methodological rather than astrophysical. Future high-cadence, high-precision timing observations with facilities such as FAST, SKA, and eXTP will be required to resolve glitch rise times and transient overshoots at millisecond resolution, which can, in principle, provide stringent tests of dynamical models once those models incorporate more realistic 3D multi-fluid hydrodynamics, improved treatments of vortex pinning and creep, and consistent entrainment in both crust and core.

Finally, while glitches are generally interpreted as sudden angular momentum transfer from a pinned superfluid reservoir to the crust, the opposite phenomenon—anti-glitches—has also been reported, most prominently in magnetars and tentatively in several rotation-powered pulsars. Within a dynamical framework, anti-glitches correspond to a net loss of crustal angular momentum relative to the superfluid. From the perspective of frictional coupling, both glitches and anti-glitches probe the same underlying physics, but in different regimes: glitches constrain the efficiency and timescale of vortex unpinning and outward angular momentum transfer, whereas anti-glitches reveal conditions under which angular momentum is withheld from the crust or external braking is enhanced. Studying glitches and anti-glitches within a unified dynamical framework therefore expands the accessible parameter space of superfluid friction and provides complementary sensitivity to the EOS, especially when combined with coordinated multi-wavelength monitoring.

\section*{Acknowledgments}
We are thankful to the XMU neutron star group for helpful discussions.
The work is supported by the National Natural Science Foundation of China (grant Nos. 12273028, 12494572).

\bibliography{friction}{}
\bibliographystyle{aasjournal}

\end{document}